\documentstyle[epsf,twocolumn,pre,aps,amssymb]{revtex}
\begin{document} 
\wideabs{    
\draft
\title{ Orientational orders in binary mixtures of hard HGO molecules} 
\author{ Xin Zhou$^1$, Hu Chen$^2$ and Mitsumasa Iwamoto$^1$}
\address{$\ ^{1}$Department of Physical Electronics, Tokyo Institute of 
Technology, O-okayama 2-12-1, Meguro-ku, Tokyo 152-8552, Japan \\
$\ ^{2}$Department of Civil Engineering, National Univeristy of Singapore,
10 Kent Ridge Crescent, Singapore 119260}
\date{\today}

\maketitle

\begin{abstract}
Based on a standard constant-NPT Monte Carlo molecular simulation, we have 
studied liquid crystal phases of binary mixtures of non-spherical molecules. 
The components of the mixtures are two kinds of hard Gaussian overlap (HGO) 
molecules, one kind of molecules with a small molecular-elongation parameter 
(small HGO molecules) cannot form stable liquid crystal phase in bulk, and 
other with a large elongation parameter (large HGO molecules) can form liquid 
crystal phase easily. 
In the mixtures, like the large HGO molecules, the small HGO molecules can also
form an orientation-ordered phase, which is because that the large HGO 
molecules can form complex confining surfaces to induce the alignment of the 
small molecules and generate an isotropic-anisotropic phase transition in  
the whole binary mixtures. 
We also study the transition on different mixtures composed of small and large 
HGO molecules with different elongations and different concentrations 
of the large molecules. 
The obtained result implies that small anisotropic molecules might show 
liquid crystal behavior in confinement. 
\end{abstract}
}

\newpage
\section{Introduction}
Surface-induced ordering of liquid crystal (LC) molecules is of important 
technological and scientific interests~\cite{Haaren2001,Chaudhari2001}. 
Very recently, Boamfa~\cite{Boamfa2003} {\it et al.} experimentally observed 
an isotropic-nematic (IN) surface phase transition in a mixture of nematic 
LCs on a substrate. Usually, in confined LCs, interaction between 
the LC molecules and the surfaces of the confining walls gives rise to a 
nematic phase at the surface, though the bulk is still an isotropic 
phase~\cite{Boamfa2003,Nakanishi1982,Bonn2001}. Based on Monte-Carlo (MC) 
simulation methods, Gruhn and Schoen~\cite{Gruhn1997,Gruhn1998} have studied 
the microscopic structure of molecularly thin confined LC films. They found 
that the orientationally ordered thin LC films were in thermodynamically 
equilibrium with the isotropic bulk phase. 
In our recent work~\cite{Zhou2003}, based on MC simulations of 
ellipsoid-like molecules confined in slit pores, we found a similar 
phenomenon in confined LC molecules. Additionally, as ellipsoid molecules 
with a small molecular-elongation parameter are confined in very thin slit 
pores, the surface of the pores can induce an orientation-ordered phase 
which cannot be formed in bulk due to the very small 
elongation parameter of the studied molecules. The result implies that the 
surface of confining walls not only induces surface IN phase transition in LC 
materials, but also stabilizes the orientation-ordered phases in non-LC 
materials (in the bulk, non-LC materials such as 
small-molecular-weight fluids cannot form aniostropic phases). 
Since small-weight molecules can response very fast to the variation of 
external fields, the small-molecular-weight LCs 
would be of great advantage in many applications. However, in simple 
confined geometry, the surface-induced effects are limited to the several 
molecular layers neighboring the surfaces of the confining 
walls~\cite{Boamfa2003,Zhou2003}, but in complex confined geometry, we may 
expect surface-induced effects in whole fluids. 

Polymer-dispersed liquids can be used to study the properties of fluids 
confined in complex geometry, since the dispersed polymers form complex 
networks. Recently, Chiccoli {\it et al.} studied the 
network-induced ordering in LCs based on MC simulations~\cite{Chiccoli2002}.
In the mixture of polymers and LC molecules, the polymers are thought as 
``surfaces" to confine the LC molecules and the surface effect exists in the 
whole system. We may address the following question: Do the found 
properties near the confining walls in simple confined geometry globally exist 
in the complex confined geometry (i.-e. mixtures)? However, we are interested
in a particular question in this paper: Since small ellipsoid-like molecules 
(no orientation-order phases in bulk) can be induced an IN transition in a 
thin slit pore~\cite{Zhou2003}, can we find similar results in complex 
pores? For example, in mixtures of the small ellipsoid-like molecules and 
other molecules, can IN transition occur? 
We will consider a binary mixture composed of small ellipsoid molecules 
and usual LC molecules. These LC molecules can form complex surfaces to 
affect the small molecules. Can the surfaces induce ordering phases in the 
small (non-LC) molecules? 
Since there are large amount of small-weight molecules in the 
mixture, the systems may show some characteristics of expected 
small-molecule LCs.

In this paper, based on standard constant-pressure (NPT-ensemble) Monte Carlo 
(MC) simulations, we study binary mixtures of hard Gaussian overlap (HGO) 
molecules. One component of the mixtures is large HGO molecules with large 
molecular elongation parameter $k$ (defined as the length-to-breadth ratio), 
and other is small HGO molecules with small $k$. The large HGO molecules 
mimic LC molecules which can form LC phases in bulk alone, whereas the small 
HGO moleucles, corresponding to normal small anisotropic molecules, cannot 
form stable LC 
phases in bulk. Hence each mixture is a solution with LC solute and non-LC 
solvent. Mixtures of LC molecules and other molecules have been studied, such 
as earlier researches of LCs with small amounts of 
impurities~\cite{McColl1975,Dubault1980,Sigaud1986} and recent studies 
on mixture of polymers and LCs~\cite{Chiccoli2002,Roussel2003}. But few works
focused on the ordered phases of the solution of LCs in small molecules.
Here, we calculate the equation of states and orientational ordering 
parameter of the mixtures. 
Similar to our previous finding that confining walls induce an alignment of 
small HGO molecules in thin slit pores~\cite{Zhou2003}, we find, 
in the mixture, the large-$k$ (LC) molecules can induce an alignment of the 
small-$k$ (non-LC) molecules, hence form an orientation-ordered phase in the 
whole system. 
Our results indicate that small molecules will form LCs with the help of 
LC molecules in their mixtures. 

The paper is organized as follows: In Sec. II, we describe a HGO model and 
give an explicit formula of intermolecular interaction between different-$k$ 
HGO molecules, then we 
describe the presentation of our simulations. Simulation results are shown 
in Sec. III. Finally we conclude and discuss in Sec. IV. 


\section{Potential Model and Computational Details}
There is a great number of molecular simulations in the literature on the 
properties of LCs. Among these simulations, LC 
molecules are described using anisotropic interactive 
models~\cite{Gruhn1998,Miguel2003,Miguel2001,Padilla1997,Dijkstra2001,McGrother1996,Veerman1990,Venkat1987,Miguel2002,Miguel2002b,Gay1981}. Although the 
detail results depend on the models used, some general conclusions have been 
found: Repulsive interactions between molecules play key roles to the LC 
structures, so some hard nonspherical models~\cite{Miguel2003,Miguel2001,Padilla1997,Dijkstra2001,McGrother1996,Veerman1990,Venkat1987} have been widely 
employed to study the properties of LCs; LC phases can form only when the 
anisotropic parameters of molecular interactions are greater than the 
critical values. Otherwise (i.e. for molecules with small 
anisotropic interactive parameters), as one increases the pressure or 
decreases the temperature, these systems will freeze or crystallize before 
forming LC phases~\cite{Zhou2003,Rigby1989}. Many recent works focus on the 
macroscopic properties and the phase transitions of models with large 
anisotropic parameters~\cite{Gruhn1997,Gruhn1998,McGrother1996,Veerman1990,Miguel2002,Lange2002}.

Among LC models, the HGO 
model~\cite{Zhou2003,Miguel2003,Miguel2001,Padilla1997,Rigby1989} is widely 
used. The model came from the Gaussian overlap potentials developed by Berne 
and Pechukas~\cite{Berne1972}. It is supposed that the mass (or electron) 
density of molecules was a Gaussian function of space vector $\mathbf x$,
\begin{eqnarray}
  G_{i}({\mathbf x}) = \exp (- {\mathbf x} \cdot {\mathbf A}_{i}^{-1} 
  \cdot {\mathbf x}), \nonumber \\
  \mathbf{A} = ({\sigma}_{\parallel}^{2} - {\sigma}_{\perp}^{2}) 
  {\mathbf u}_{i} {\mathbf u}_{i} + {\sigma}_{\perp}^{2} {\mathbf I}.
\end{eqnarray}
Here ${\mathbf u}_{i}$ is a unit vector along the 
principal axis of the $i$th molecule and ${\mathbf I}$ is the unit matrix, 
$\sigma_{\parallel}$ and $\sigma_{\perp}$ correspond to the size of the 
molecules along and perpendicular to its principal axis, respectively. 
The interaction between two molecules $i$ and $j$ is supposed to be 
proportional to the overlap of the two molecules, 
\begin{eqnarray}
V({\mathbf u}_{i}, {\mathbf u}_{j}, {\mathbf r}_{ij}) &\sim& {\mid} {\mathbf 
A}_{i} {\mid}^{-1/2} {\mid} {\mathbf A}_{j} {\mid}^{-1/2} \int 
d {\mathbf x} G_{i}({\mathbf x}) G_{j}({\mathbf x}-{\mathbf r}_{ij}) 
\nonumber \\
&\sim& {\mid} {\mathbf A}_{i} + {\mathbf A}_{j} {\mid}^{-1/2} 
{\exp} [- {\mathbf r} \cdot ({\mathbf A}_{i} + {\mathbf A}_{j})^{-1}
\cdot {\mathbf r}] \nonumber \\
&\sim& \epsilon({\mathbf u}_{i}, {\mathbf u}_{j}) 
\exp [ - r^{2}/{\sigma}^{2}({\mathbf u}_{i}, {\mathbf u}_{j}, 
{\hat r})], 
\end{eqnarray}
where, $r$ and ${\hat r}$ are the length and the unit vector of the vector 
${\mathbf r}_{ij}$, respectively. We have, 
\begin{eqnarray}
{\sigma}^{-2}({\mathbf u}_{i}, {\mathbf u}_{j}, {\hat r}) = 
{\hat r} \cdot ({\mathbf A}_{i} + {\mathbf A}_{j})^{-1} 
\cdot {\hat r}.
\end{eqnarray}
If the three eigenvalues and corresponding unit eigenvectors of 
${\mathbf A}_{i} + {\mathbf A}_{j}$ are
noted as $\lambda_{k}$ and ${\hat v}_{k}$, ($k=1,2,3$), respectively, we have
\begin{eqnarray}
{\sigma}^{-2}({\mathbf u}_{i}, {\mathbf u}_{j}, {\hat r}) = \sum_{k=1}^{3} 
({\hat r} \cdot {\hat v}_{k})^{2}/\lambda_{k}.
\end{eqnarray}
Actually, since a vector perpendicular to ${\mathbf u}_{i}$ and 
${\mathbf u}_{j}$ is an eigenvector of 
${\mathbf A}_{i} + {\mathbf A}_{j}$, we only need to diagonalize 
a $2 \times 2$ matrix. Since ${\hat v}_{k}$ ($k=1,2,3$) are unit vectors 
and perpendicular each other, we have $\sum_{k=1}^{3} 
({\hat r} \cdot {\hat v}_{k})^{2} = 1$. Using these relationships, we easily 
obtain,
\begin{eqnarray}
\sigma({\mathbf u}_{i}, {\mathbf u}_{j}, {\hat r}) = \sigma_{0} 
\{1- \frac{{\cal X}_{ij}}{2} [\frac{2(1+\theta)}{1+
{\cal X}_{ij} \theta} ({\hat r} \cdot {\hat v}_{1})^{2} + 
\frac{2(1-\theta)}{1 - {\cal X}_{ij} \theta} ({\hat r} \cdot 
{\hat v}_{2})^{2} ] \}^{-1/2}, 
\label{sigmaofhgo}
\end{eqnarray}
where $\sigma_{0}=\sqrt{\sigma_{i \perp}^{2} + \sigma_{j \perp}^{2}}$, and
$\theta= \sqrt{ {\alpha}_{ij}^{2} + (1-{\alpha}_{ij}^{2}) 
({\mathbf u}_{i} \cdot {\mathbf u}_{j})^{2} }$. ${\cal X}_{ij}$ and 
$\alpha_{ij}$ are writen as,
\begin{eqnarray}
{\cal X}_{ij} &=& \frac{ ({\sigma}_{i \parallel}^{2} - {\sigma}_{i \perp}^{2}) 
 + ({\sigma}_{j \parallel}^{2} - {\sigma}_{j \perp}^{2}) } 
{ ({\sigma}_{i \parallel}^{2} + {\sigma}_{i \perp}^{2}) + 
({\sigma}_{j \parallel}^{2} + {\sigma}_{j \perp}^{2}) }, 
\nonumber \\
{\alpha}_{ij} &=& \frac{ ({\sigma}_{i \parallel}^{2} - {\sigma}_{i \perp}^{2}) 
 - ({\sigma}_{j \parallel}^{2} - {\sigma}_{j \perp}^{2}) }
{ ({\sigma}_{i \parallel}^{2} - {\sigma}_{i \perp}^{2}) +
({\sigma}_{j \parallel}^{2} - {\sigma}_{j \perp}^{2}) }, 
\label{Size} 
\end{eqnarray}
respectively. 
The directions of the unit eigenvectors ${\hat v}_{1}$ and ${\hat v}_{2}$
are given as,
\begin{eqnarray}
{\vec v}_{1} = (\theta + \alpha_{ij}) {\mathbf u}_{i} + 
({\mathbf u}_{i} \cdot {\mathbf u}_{j}) (1-\alpha_{ij}) {\mathbf u}_{j}, \\
{\vec v}_{2} = ({\mathbf u}_{i} \cdot {\mathbf u}_{j}) (1+\alpha_{ij}) 
{\mathbf u}_{i} - (\theta + \alpha_{ij}) {\mathbf u}_{i},
\end{eqnarray}
respectively. Obviously, if the $i$th molecule and the $j$th molecule are 
identical, we obtain the usual formula of identical 
molecules~\cite{Berne1972},
\begin{eqnarray}
\sigma({\mathbf u}_{i}, {\mathbf u}_{j},{\hat r}, ) = \sigma_{0} 
\{ 1 - \frac{\cal X}{2} [ \frac{ ({\hat r} \cdot {\mathbf u}_{i} + 
{\hat r} \cdot {\mathbf u}_{j})^{2}} {1 + {\cal X} 
({\mathbf u}_{i} \cdot {\mathbf u}_{j})} + 
\frac{ ({\hat r} \cdot {\mathbf u}_{i} - 
{\hat r} \cdot {\mathbf u}_{j})^{2}} {1 - {\cal X} 
({\mathbf u}_{i} \cdot {\mathbf u}_{j})} ] \}^{-1/2}, 
\end{eqnarray} 
where ${\cal X}_{ij}$ is simply noted as ${\cal X}$.

The interactive potential in HGO model is simplified as, 
\begin{eqnarray}
V({\mathbf u}_{i}, {\mathbf u}_{j},{\mathbf r}_{ij}) = \left\{ 
\begin{array}{@{\,}ll}
 \infty, & {\rm if} \ \mbox{$ r_{ij} \le \sigma({\mathbf u}_{i}, 
{\mathbf u}_{j}, {\hat r}_{ij}) $} \\
0, & {\rm if} \ \mbox{$ r_{ij} > \sigma( {\mathbf u}_{i}, 
{\mathbf u}_{j}, {\hat r}_{ij}) $}.
\end{array}
\right.
\end{eqnarray}
where $\sigma({\mathbf u}_{i}, {\mathbf u}_{j}, {\hat r}_{ij})$ is defined as
Eq. (\ref{sigmaofhgo}). The HGO model is thought as an approximation for the 
excluded volume of hard ellipsoids of revolution~\cite{Padilla1997,Rigby1989}.
In this paper, we study binary mixtures consisting of two kinds of HGO 
molecules with molecular elongation $k_{L}$ and $k_{S}$, respectively. Here 
the elongation $k$ is defined as,
\begin{eqnarray} 
k_{i}= {\sigma}_{i \parallel}/{\sigma}_{i \perp}.
\end{eqnarray}
Here $i= L$ or $S$, corresond to large and small HGO molecules, respectively. 
For simplification, we suppose $\sigma_{L\perp} = \sigma_{S\perp}$, 
then ${\sigma}_{0}$ is a constant which is set as unit of length in our 
simulations, and 
${\cal X}_{ij}$ can be writen as 
\begin{eqnarray}
{\cal X}_{ij} = (k_{i}^{2}-1 + k_{j}^{2}-1)/(k_{i}^{2}+1 + k_{j}^{2}+1).
\end{eqnarray}
 
We use standard constant-pressure (const-NPT) MC simulations to 
obtain the orientational orders and the equation of state $P(\rho)$ 
( where $P$ corresponds to the real 
$P \sigma_{0}^{3}/k_{B}T$). The simulations are performed with $N=500$ or 
$256$ molecules which consist of $N_{L}$ large-elongation molecules and 
$N - N_{L}$ small-elongation molecules. The selected $k_{S}$ is 
about $2.0$ and $k_{L}$ is between $3.0$ and $6.0$. 
According to a large number of previous 
simulations~\cite{Zhou2003,Miguel2003,Miguel2001,Padilla1997,Rigby1989}, 
the selected large-$k$ molecules can form 
orientation-ordered phases in bulk alone, but the selected small-$k$ 
molecules cannot form stable LC phases in bulk up to 
a typical high density of liquids~\cite{Zhou2003,Rigby1989}. 
In our simulations of the mixtures, the simulating box is cubic and equally 
fluctuates in three directions, cubic periodic boundary conditions are used. 
The simulations are organized in MC cycles, 
each MC cycle consisting (on average) of $N$ trial translational and 
rotational molecular displacements and one trial volume fluctuation. The trial
displacements of two kinds of molecules are separately treated. The 
maximum step length of each trial move is automatically chosen at each 
pressure for making the acceptable probability fall between $0.4$ and $0.5$. 
The starting configuration is a face-center-cubic lattice which melt at 
low pressure and equilibrate for $10^{5}$ MC cycles. The 
system is slowly 
compressed in small pressure steps. For any given pressure, the system is 
typically equilibrated for $10^{5}$ MC cycles and an average 
is taken over all configurations of additional $10^{5}$ MC 
cycles, then the final 
configuration is set as the starting configuration of the next pressure. 
At some higher pressure (density) zones, $3 \sim 5$-times longer MC cycles 
are used in equilibrated and averaged processes. 
 
The orientational order parameter $S$ is calculated in the simulations as 
the largest eigenvalue of the ordering $Q_{\alpha \beta}$ tensor, defined 
in terms of the components of the unit vector $u_{i \alpha}$ along the 
principal axis of the molecules,
\begin{eqnarray}
Q_{\alpha \beta} = \frac{1}{N} \sum_{i=1}^{N} ( {3 \over 2} u_{i \alpha} 
u_{i \beta} - {1 \over 2} \delta_{\alpha \beta} )
\label{orderparameter}
\end{eqnarray}
Besides detecting the order parameter of all molecules, we 
also separately calculate the order parameter ($S_{L}$ and $S_{S}$) of 
larger and smaller molecules by limiting the average of 
all molecules in Eq. (\ref{orderparameter}) to single kind of molecules.

\section{Results}
\begin{figure}
\centerline{\epsfxsize=10cm \epsfbox{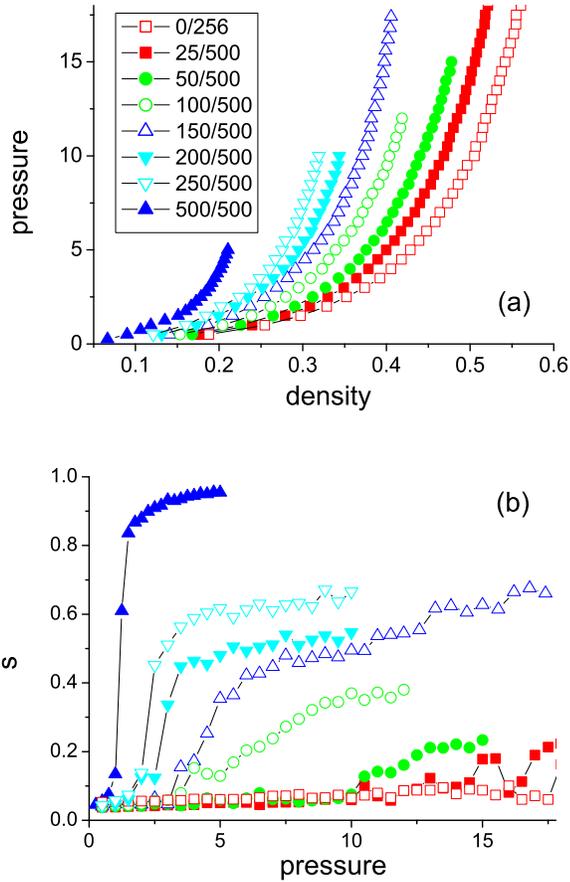}}
\caption{ (a) The Equation of state (pressure versus density) of binary 
mixtures with two kinds of HGO molecules. Their molecular elongation $k$ are 
$5.0$ and $2.0$, respectively. The data in label correspond to $N_{L}/N$ of 
each curves. For $N_{L}=0$, we select $N = 256$. 
(b) The Orientation order parameter $S$ versus pressure of the mixtures. All 
parameters and symbols are same as that in Fig. \ref{fig1}. It is clearly 
shown an Isotropic-nematic phase transition except in small $x$ cases. 
However, except a trivial case ($x=500/500=1.0$), in nematic phase, the order 
parameter $S$ of the mixture does not arrive at unit.
\label{fig1}}
\end{figure}

We first simulate mixtures composed of $N_{L}$ large HGO molecules 
($k_{L}=5$) and $N_{S} = N - N_{L}$ small HGO molecules ($k_{S}=2.0$). 
The concentration of large molecules is $x = N_{L}/N$. We present the 
equation of 
state and orientation order parameter $S$ of the mixtures with different $x$
in Fig. \ref{fig1}. In label of the figures, $N_{L}/N$ is listed. For pure 
small HGO molecules($k=2.0$), the 
simulations are performed with $N=256$ molecules, but for other cases, we 
simulate systems with $N=500$ molecules. In Fig. \ref{fig1} (a), the equations 
of state exhibit a discontinuity for all larger-$x$ systems. From 
Fig. \ref{fig1}(b), we can clearly find the spontaneous phase transition 
between isotropic and anisotropic phases in the HGO system conposed 
of small amounts of large-$k$ molecules. However, the order parameter 
$S$ of the mixture is smaller than unity, which implies that the molecules 
only partially align or only partial molecules align. With decreasing 
concentration of large molecules $x$, the transition pressure increases 
and the corresponding order 
parameter $S$ in the anisotropic phase decreases.
 
\begin{figure}
\centerline{\epsfxsize=10cm \epsfbox{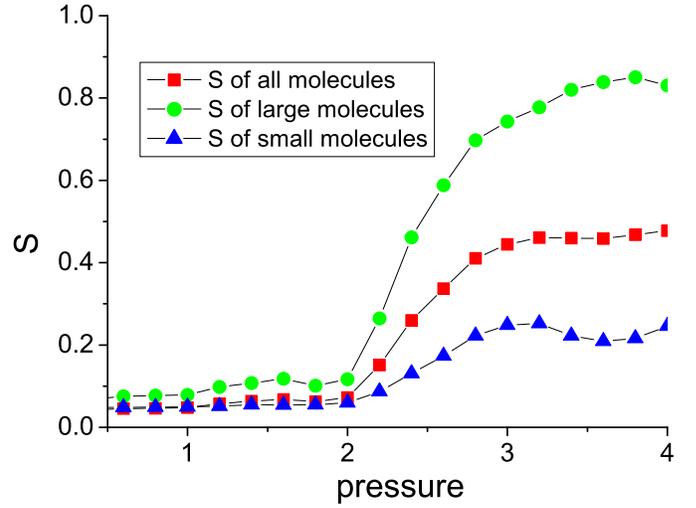}}
\caption{  Orientation order parameters of large, small and all molecules 
versus pressure $P$ in a mixtures. Where $x=200/500$, $k_{S}=2.2$ and 
$k_{L}=5.0$.  
\label{fig2}}
\end{figure}

To detect the microscopic structure of two kinds of molecules in the 
mixtures, we calculate separately 
the orientation order parameter of large and small HGO molecules. 
In Fig. \ref{fig2}, we show relations between the three order parameters $S$ 
and the pressure in a mixture, where $x=0.4$, $k_{S}=2.2$ and $k_{L}=5.0$. 
These order parameters describe the alignment of large-$k$ molecules, 
small-$k$ molecules, and all molecules, respectively. From Fig. \ref{fig2}, we 
find a transition for both large and small molecules beginning at a same 
pressure $P \sim 2.0$ in the mixture. With increasing pressure, the large 
molecules almost completely align, but the small molecules only partial 
align. Thus the average S of all molecules is far smaller than unity. 
We also find the direction of the large molecules is parallel to that of the 
small molecules (not shown). Therefore, it can be concluded that the alignment 
of the small molecules is unambiguously ascribed to the alignment of of the 
large molecules. Additionally, the transition seems to occur at a pressure 
range, rather than at an exact value, which may correspond to the known I-N 
coexistence in LC mixtures.
In Fig. \ref{fig3}, we show the orientation orders of large molecules and 
small molecules with different $k_{L}$ and $k_{S}$ values. The IN 
transition value of small molecules nematic order parameter decreases as the 
elongation of large molecules $k_{L}$ increases. This surprising result is 
ascribed to the lower transition pressure as $k_{L}$ is larger 
(see Fig. \ref{fig3}(d) ).

\section{Concluding Remarks}
From a great number of researches of LCs, people had known that only 
molecules with larger 
anisotropic parameters can form LC phases in bulk. For example, the molecular 
elongation of normal LC molecules is larger than $4$. In our previous 
work~\cite{Zhou2003}, we found that small anisotropic molecules may form 
stable LC phases as the molecules are confined in very thin pores. The result
indicates that some normal inorganic small molecules (whose $k$ is about $2$) 
may form LC phases in 
confined geometries (or near surfaces). In current widely studied 
polymer-dispersed LC systems, the polymers can be thought as complex surfaces 
confining LC molecules, so we guess that small anisotropic molecules may 
also show LC phases due to the existing large (LC or polymer) 
molecules in the mixtures. 

In this paper, we present results from standard NPT-ensemble simulations of 
mixtures of small and large HGO molecules. Among the two components, the small 
HGO molecules cannot form LC phases alone in bulk, but large HGO molecules 
can. We detect possible isotropic-anisotropic
phase transition in the mixtures. The results presented here give clear 
evidence of spontaneous formation of orientation-order phases in the mixtures.
We find that the spontaneous ordering of the large and small molecules
occur at the same pressure. It implies that the ordering of the small 
molecules is ascribed to the inducement of the large molecules. We also study 
the 
induced phase transition behaviour for different large-molecule concentrations
and different molecular elongations of the large and small molecules.

The confinement-induced LC phases in mixtures present here may be useful. In 
the mixtures, the viscosity should be smaller than large-molecule systems 
(usual LCs), so we think that both the large and the small molecules may 
faster align in external field, which implies that the mixtures may show 
the characteristics of expected small-molecule LCs, such as short response 
time. However, it is necessary to study the confined effects based on more 
reliable molecular models.

X. Z is financially supported by the Grants-in-Aid for 
Scientific Research of JSPS; H. Chen is supported by the Singapore 
Millennium Scholarship.  

(Correspondent address: zhou@pe.titech.ac.jp).

\begin{figure}
\caption{ Orientation order parameters of large and small molecules versus 
pressure in mixtures with different $k_{S}/k_{L}$ for detecting the 
elongational dependence of the ordering parameters. For systems with 
$k_{L}=3.0$, $N_{L}/N = 101/256$; for other systems, $N_{L}/N =200/500$.
The symbols of (b) are same as that of (a) and the symbols of (c) are same
as that of (d). (a) Orientational order parameter $S_{L}$ of large molecules 
in mixtures versus $P$ (only near transition zone); (b) Orientational order 
parameter $S_{S}$ of small molecules in mixtures versus $P$; (c) $S_{L}$ 
versus $P$; (d) $S_{S}$ versus $P$. In (c) and (d), for the case of 
$k_{S}/k_{L}=2.0/5.0$, we only show data in lower pressure range (triangle).
\label{fig3}}
\end{figure} 
 
\end{document}